\documentclass[10pt]{article}
\usepackage{graphicx}
\usepackage{times}
 at 14truept
 at 10.0truept
\title{PIPELINE PROCESSING OF VLBI DATA}
\author{Cormac Reynolds$^{(1)}$, Zsolt Paragi$^{(1)}$, Mike Garrett$^{(1)}$\\
      $^{(1)}$Joint Institute for VLBI in Europe, Postbus 2, 7990 AA Dwingeloo,
      The Netherlands.}
\def\h0{km s$^{-1}$ Mpc$^{-1}$}

\begin{document}

\pagestyle{empty}
\renewcommand{\baselinestretch}{2.0}
\linespread{2.0}
\normalsize

\begin{centering}
\textbf{
{\fontsize{12}{0}\selectfont PIPELINE PROCESSING OF VLBI DATA}\\
\underline{Cormac Reynolds}$^{(1)}$, Zsolt Paragi$^{(2)}$, Mike Garrett$^{(3)}$\\
\textit{$^{(1)}$\underline{Joint Institute for VLBI in Europe, Postbus 2, 7990 AA Dwingeloo,
The Netherlands. E-mail: reynolds@jive.nl} \\
$^{(2)}$As (1) above, but E-mail: zparagi@jive.nl\\
$^{(3)}$As (1) above, but E-mail: garrett@jive.nl
}
}

\end{centering}

\renewcommand{\baselinestretch}{1.0}
\linespread{1.0}
\normalsize

\section*{ABSTRACT}

As part of an on-going effort to simplify the data analysis path for VLBI
experiments, a pipeline procedure has been developed at JIVE to carry out much
of the data reduction required for EVN experiments in an automated fashion.
This pipeline procedure runs entirely within AIPS, the standard data reduction
package used in astronomical VLBI, and is used to provide preliminary
calibration of EVN experiments correlated at the EVN MkIV data processor. As
well as simplifying the analysis for EVN users, the pipeline reduces the delay
in providing information on the data quality to participating telescopes, hence
improving the overall performance of the array. A description of this pipeline
is presented here.

\section*{INTRODUCTION}

At the current time Very Long Baseline Interferometry (VLBI) datasets require
considerable effort and detailed knowledge on the part of the astronomer to
produce a calibrated dataset suitable for scientific analysis. This places high
manpower requirements on any projects involving large amounts of VLBI data,
such as surveys and monitoring campaigns. The high data rates and the
requirement for real-time processing of future radio interferometers (e.g.
ALMA, LOFAR, SKA) will create a situation where traditional reduction data
methods will no longer be viable, necessitating a large element of automatic
data reduction.

As a result, efforts are being made to produce `pipelines' which will perform
much of the data analysis in an automated way. The ultimate goal of these
pipelines would be to provide a fully calibrated dataset to the astronomer,
where the quality of the calibration is sufficient to meet the scientific aims
of the experiment. 

As a first step towards this goal, a pipeline procedure has been developed at
JIVE to calibrate the regular Network Monitoring Experiments (NMEs) carried out
by the European VLBI Network (EVN), and to perform the initial calibration of
EVN user experiments. The NMEs are simple experiments, typically observing two
sources in a continuum, phase-referencing mode and are intended to give an
indication of the quality of the data and calibration information provided by
the various stations making up the array.  The simple nature of the NMEs makes
them ideal candidates for pipeline processing.  As an extension to the NME
analysis, the pipeline can also be used to process data on calibrator scans in
user experiments in order to provide more frequent feedback on the performance
of the network than is provided by the NMEs alone.  Finally, the pipeline
procedure, since it carries out all of the main calibration tasks required in a
typical VLBI experiment, can be used as the basis to provide a simplified data
reduction path for many user experiments.  This does not yet provide a truly
automated reduction package as some data inspection and manipulation is still
required on the part of the astronomer.  The aim is to minimise the time and
effort required of the astronomer in performing tasks that can be accomplished
by the pipeline.

\section*{DESCRIPTION OF THE PIPELINE}

The EVN pipeline has been written as a procedure within AIPS (produced and
distributed by NRAO); this is the standard software package used for
calibration of astronomical VLBI data. The AIPS package provides a suite of
tasks to perform the various steps of data calibration, as well as tasks that
allow inspection of the data and removal of corrupted data. The EVN pipeline
uses the facilities within AIPS to calibrate the dataset with a minimum of user
input.

\begin{figure}[here]
\begin{centering}
\includegraphics[width=75mm,angle=0,keepaspectratio]{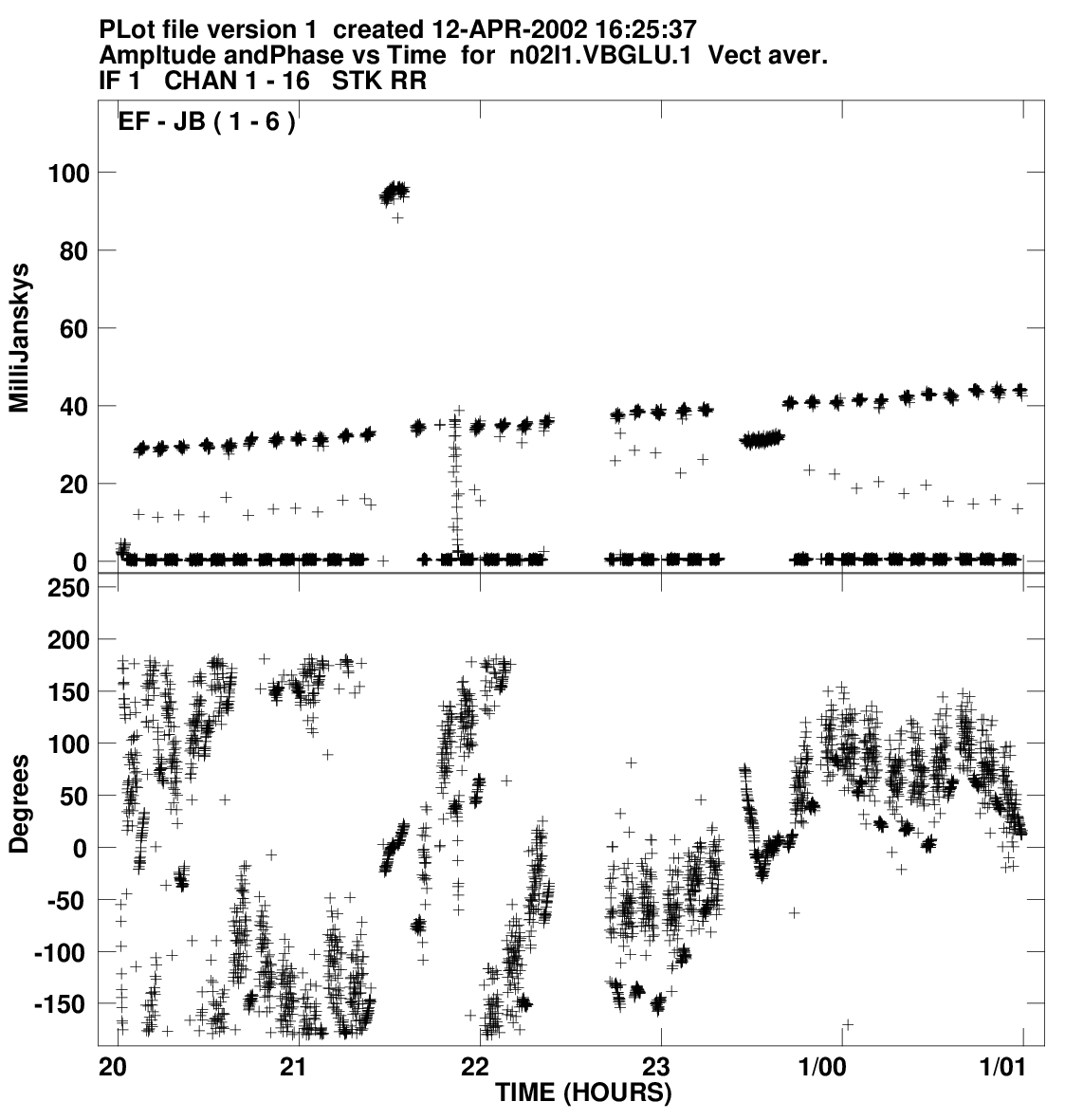}
\includegraphics[width=75mm,angle=0,keepaspectratio]{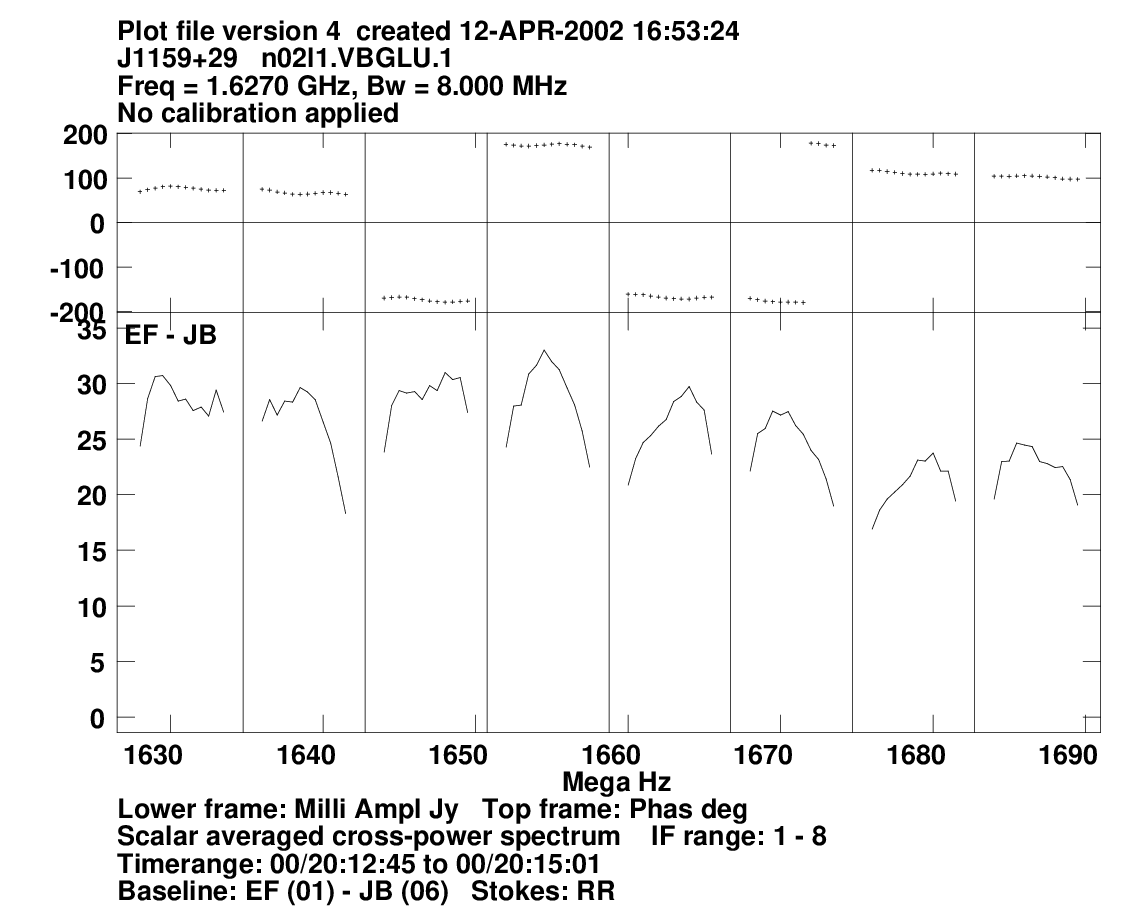}
\caption{The uncalibrated data provided by the correlator on one example
baseline; (a) the amplitude and phase of the raw data on one IF over the length
of the experiment.  (b) A scan average of the data on all the IFs (RCP only)
showing amplitude and phase as a function of observing frequency.
\label{fig:uncal}}
\end{centering}
\end{figure}

\begin{figure}[here]
\begin{centering}
\includegraphics[width=85mm,angle=0,keepaspectratio]{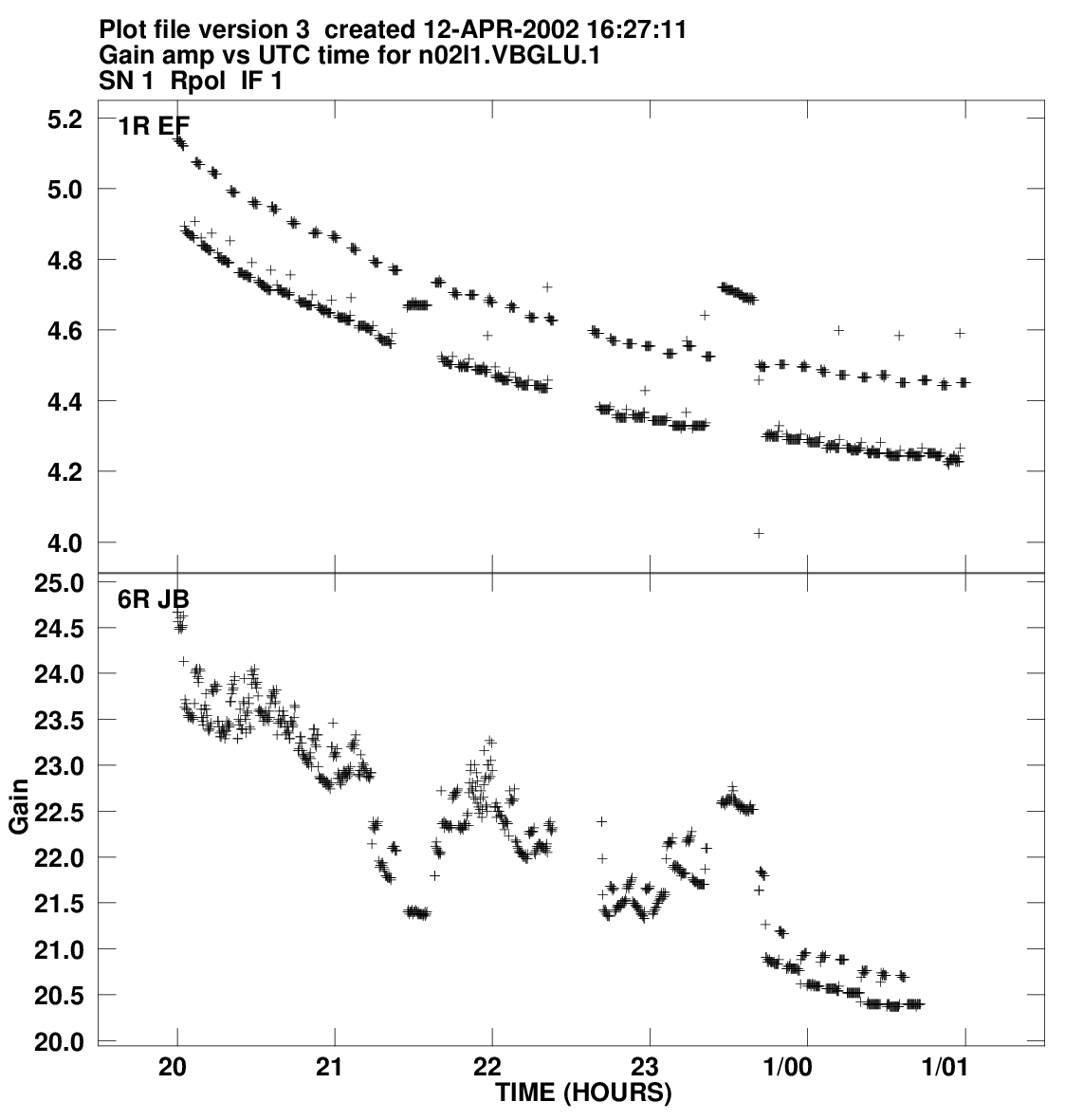}
\caption{The a priori amplitude calibration data (before interpolation) for
the example baseline. This is produced by combining a known gain curve for
each telescope with the T$_{sys}$ measured for each telescope during the
experiment. The `Gain' shown on the y-axis is in units of the square-root of
the telescope noise in Jy.
\label{fig:gain}}
\end{centering}
\end{figure}

\begin{figure}[here]
\begin{centering}
\includegraphics[width=65mm,angle=0,keepaspectratio]{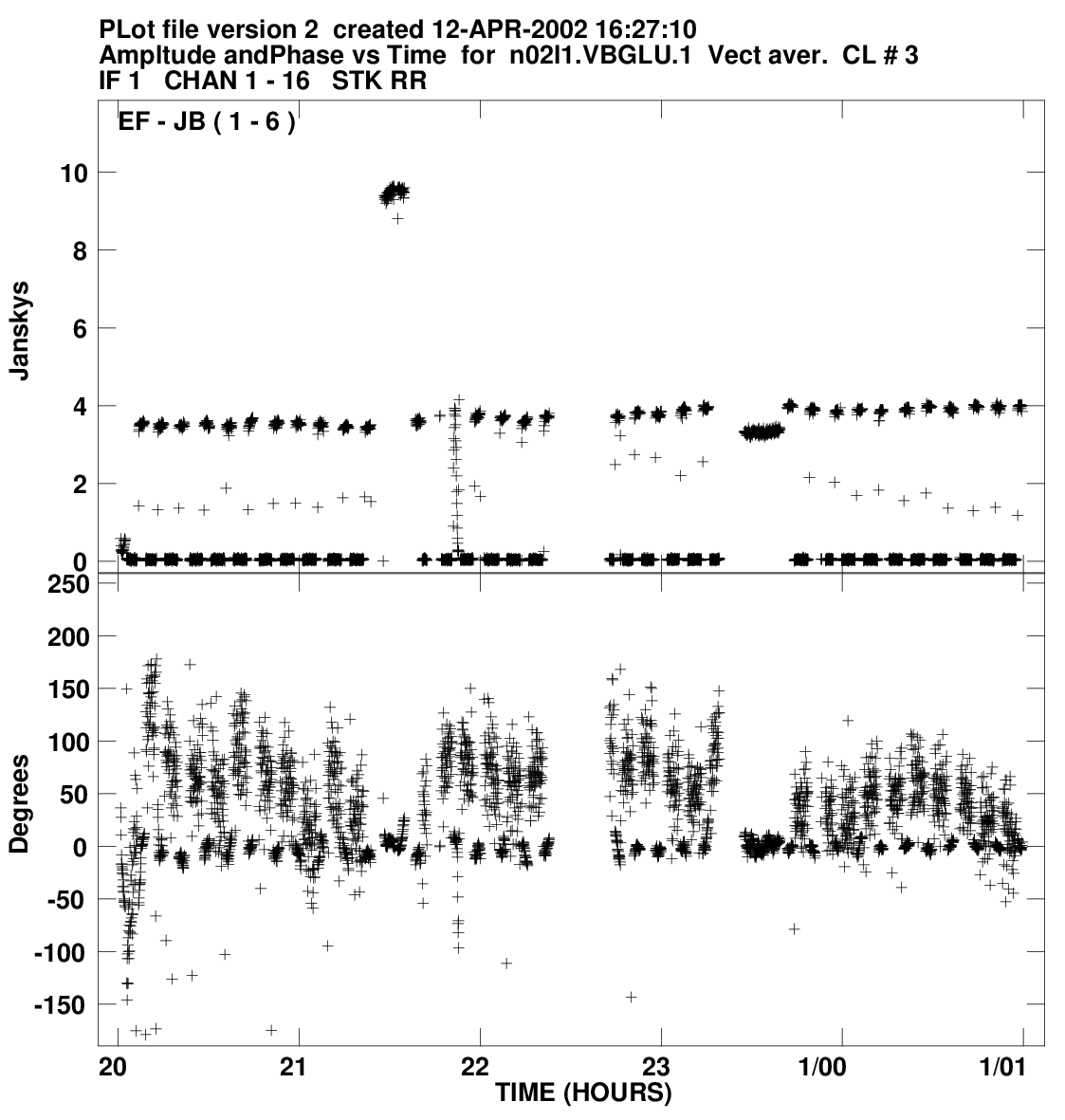}
\includegraphics[width=65mm,angle=0,keepaspectratio]{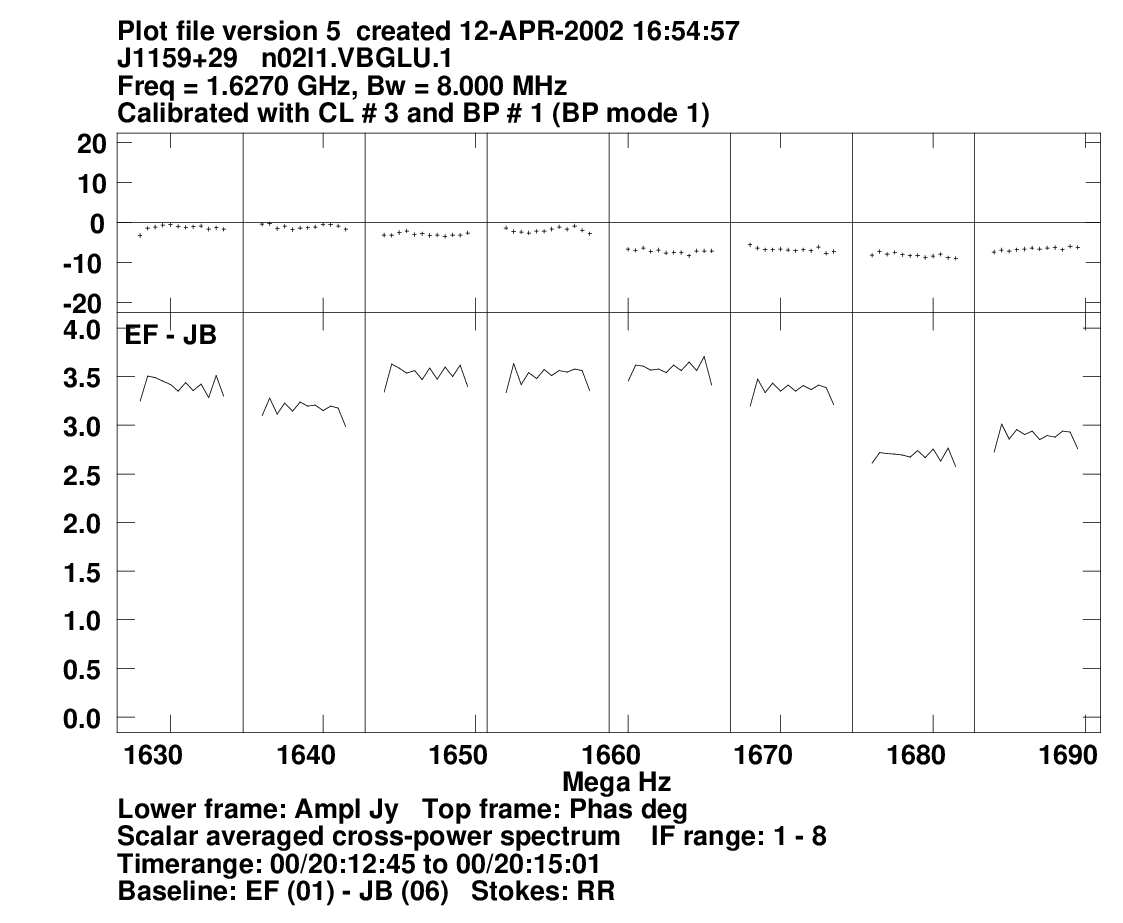}
\caption{The a priori amplitude calibrated and fringe-fitted data on the
example baseline (a) the amplitude and phase in one IF over the length of the
experiment (for both the bright phase-reference and the weak target source),
(b) A scan average of the data on all the IFs (RCP only) showing amplitude and
phase as a function of observing frequency.
\label{fig:cal}}
\end{centering}
\end{figure}

\begin{figure}[here]
\begin{centering}
\includegraphics[width=65mm,angle=0,keepaspectratio]{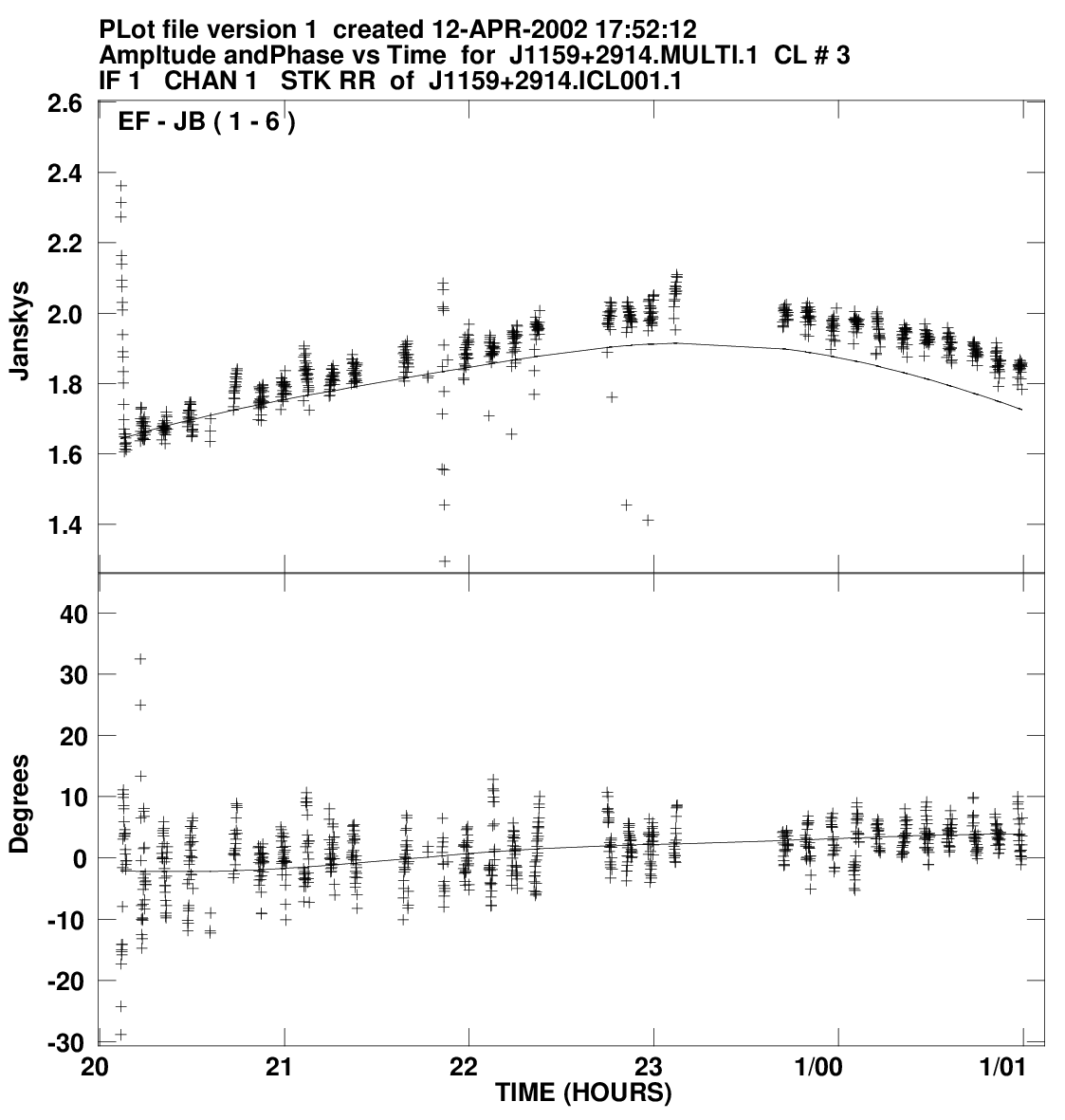}
\includegraphics[width=65mm,angle=0,keepaspectratio]{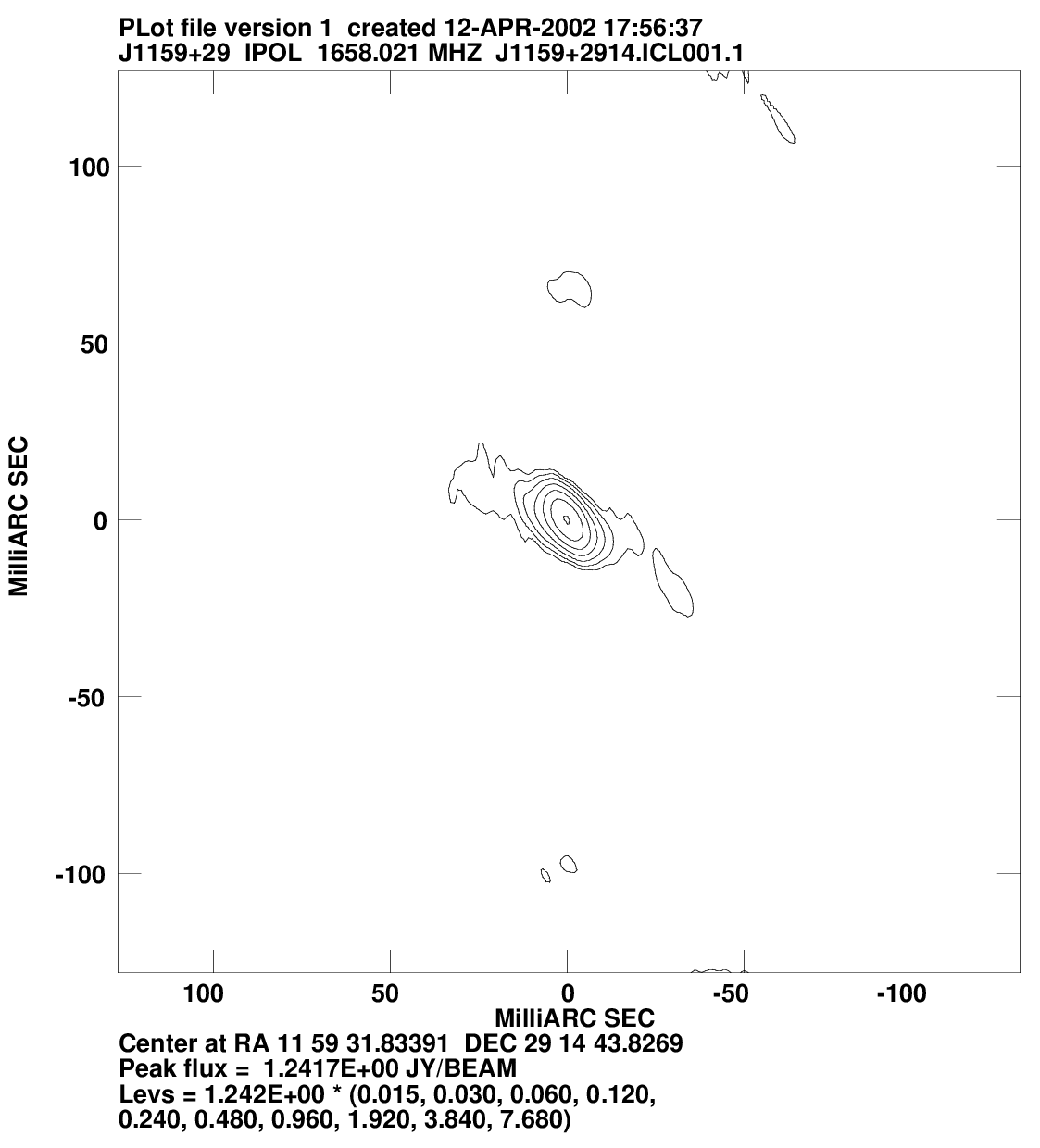}
\caption{(a) The calibrated visibilities of the calibrator source. The solid
line superposed on the visibilities represents the model of the source given by
the accompanying map. (b) The
crude map of the calibrator source produced by the pipeline after just
two iterations of self-calibration. 
\label{fig:vplot_model}}
\end{centering}
\end{figure}

A subset of the data from a recently pipelined experiment at various stages of
the pipeline procedure is presented below to illustrate the steps carried out
by the pipeline. This experiment was observed at L-band with 8 stations each
recording data at a rate of 512 Mb s$^{-1}$ [2-bit Nyquist sampling of 64 MHz
of bandwidth in two polarizations (right circular, RCP, and left circular,
 LCP)]. Two sources were observed -- a bright, compact, calibrator source, and
a weak ($\sim 10$~mJy) target source, located about 1 degree from the
calibrator. The phase-referencing method was used to calibrate the weak target
using the data from the bright calibrator.  The steps performed by the EVN
pipeline are: 

\begin{enumerate}
\item
The raw data supplied by the correlator, stored in FITS format, are loaded into
AIPS, then sorted and catalogued as required by the AIPS tasks. Data known to
be invalid from information provided by the individual stations or the
correlator is flagged. A subset of the raw data from our example experiment
is presented in Fig.~\ref{fig:uncal}.

\item
The a priori amplitude calibration is then performed using system temperatures
and gain files which are provided by the participating stations and which have
been preprocessed into a format suitable for use with the appropriate AIPS
tasks. The amplitude calibration information for our example baseline is
presented in Fig.~\ref{fig:gain}.

\item
The data are then fringe-fitted to remove frequency dependent phase slopes
across the observing band. In the case of phase-referencing experiments it is
possible to propagate solutions found for one source to another source.  
A bandpass calibration is also performed. The effects of these steps can be
seen in Fig.~\ref{fig:cal}. 

\item
Optionally, data on each baseline can then be flagged based on deviations from
the mean value for that baseline. This sort of automated flagging is only
useful for sources with simple structure.

\item
The data from any sources deemed bright enough to self-calibrate are then
iteratively self-calibrated and imaged to produce a crude map of the calibrator
sources. The calibrated visibilities and the source model, and the pipeline map
are presented in Fig.~\ref{fig:vplot_model}. After just a couple of iterations
of self-calibration reasonable agreement is found between the data and the
model, and a map with a dynamic range of $\sim 100:1$ is produced. The
solutions produced for the calibrator source can be propagated to nearby weak
target sources which are not suitable for self-calibration.

\item
At a number of stages in the calibration procedure plots of the calibrated data
are produced. These are useful to check the progress of the pipeline
calibration, to identify problems with the data and to assess the performance
of individual stations in the array.

\end{enumerate}

\section*{CONCLUSIONS}

It is intended that the majority of user experiments correlated by the EVN MkIV
data processor will be processed using the pipeline procedure. This will allow
partially calibrated datasets to be provided to users. The pipeline analysis
will also greatly increase the effectiveness of the correlator archiving policy
as the plots of the calibrator sources produced by the pipeline are made
available to all interested persons via the Internet.  Users who are interested
in archived data will have a means of instantly determining the quality of
publicly available data in the archive on an experiment by experiment basis, by
looking at the output from the pipeline.  This will make it significantly
easier to determine the usefulness of any data set for any given purpose.

The other principal benefit of the pipeline procedure is that it is now
feasible to carry out the preliminary calibration of all experiments processed
by the EVN data processor, within a very short time of the correlation being
completed. This means that the array performance (i.e. data quality and the
accuracy of the a priori calibration information) can be closely monitored
throughout observing sessions. The improved feedback to telescopes resulting
from this should result in a more reliable, and better calibrated array in the
future.

\section*{ACKNOWLEDGEMENTS}
The authors are grateful to Phil Diamond for providing the AIPS scripts 
on which the EVN pipeline is based.

\end{document}